\documentclass[aip,reprint,reqno]{revtex4-1}

\usepackage{amssymb}
\usepackage{graphicx}
\usepackage{dcolumn}
\usepackage{bm}
\usepackage{gensymb}
\usepackage[articletitle=true]{achemso}
\newcommand\Label[1]{&\refstepcounter{equation}(\theequation)\ltx@label{#1}&}

\draft

\begin{document}

\preprint{AIP/123-QED}

\title{Charging-driven coarsening and melting of a colloidal nanoparticle monolayer at an ionic liquid-vacuum interface}

\author{Connor G. Bischak}
\affiliation{Department of Chemistry, University of California, Berkeley, CA 94720.}
\affiliation{Current Address: Department of Chemistry, University of Washington, Seattle, WA 98195.}

\author{Jonathan G. Raybin}
\altaffiliation{These authors contributed equally to this work.}
\affiliation{Department of Chemistry, University of California, Berkeley, CA 94720.}

\author{Jonathon W. Kruppe}
\altaffiliation{These authors contributed equally to this work.}
\affiliation{Department of Physics, University of California, Berkeley, CA 94720.}

\author{Naomi S. Ginsberg}
\altaffiliation{Corresponding Author (nsginsberg@berkeley.edu)}
\affiliation{Department of Chemistry, University of California, Berkeley, CA 94720.}
\affiliation{Materials Sciences Division, Lawrence Berkeley National Laboratory, Berkeley, CA 94720.}
\affiliation{Department of Physics, University of California, Berkeley, CA 94720.}
\affiliation{Kavli Energy NanoScience Institute, Berkeley, CA 94720.}
\affiliation{Molecular Biophysics and Integrated Bioimaging Division, Lawrence Berkeley National Laboratory, Berkeley, CA 94720.}
\affiliation{STROBE, NSF Science \& Technology Center, Berkeley, California 94720,
United States.}

%Soft Matter instructions:
%The abstract should be a single paragraph (50–250 words) that summarises the content of the article. It will help readers to decide whether your article is of interest to them.
%It should set out briefly and clearly the main objectives and results of the work; it should give the reader a clear idea of what has been achieved. Like your title, make sure you use recognisable, searchable terms and keywords.
\begin{abstract}
We induce and investigate the coarsening and melting dynamics of an initially static nanoparticle colloidal monolayer at an ionic liquid-vacuum interface, driven by a focused, scanning electron beam. Coarsening occurs through grain interface migration and larger-scale motions such as grain rotations, often facilitated by sliding dislocations. The progressive decrease in area fraction that drives melting of the monolayer is explained using an electrowetting model whereby particles at the interface are solvated once their accumulating charge recruits sufficient counterions to subsume the particle. Subject to stochastic particle removal from the monolayer, melting is recapitulated in simulations with a Lennard-Jones potential. This new driving mechanism for colloidal systems, whose dynamical timescales we show can be controlled with the accelerating voltage, opens the possibility to manipulate particle interactions dynamically without need to vary particle intrinsic properties or surface treatments. Furthermore, the decrease in particle size availed by electron imaging  presents opportunities to observe force and time scales in a lesser-explored regime intermediate between typical colloidal and molecular systems.
\end{abstract}


\maketitle

\section{Introduction}

Colloidal physics has provided an extensive and readily accessible platform to mimic interatomic and intermolecular interactions and associated phase behavior and to discover emergent phenomena distinct from those occurring in traditional condensed matter systems.\cite{Liu_Jamming_2010,kohlstedt_self-assembly_2013, edwards_colloidal_2014,manoharan_colloidal_2015,lavergne_anomalous_2017,girard_particle_2019, soni_odd_2019,hueckel_ionic_2020} More recently, the exploration of soft matter driven away from equilibrium by various means has opened new possibilities to study and create systems in novel dynamic regimes. Irrespective of whether the individual constituents are active or passive, their dynamics are primarily visualized with optical microscopy,\cite{weeks_three_dimensional_2000,alsayed_premelting_2005,lipowsky_direct_2005,li_modes_2016,li_assembly_2016,thorneywork_two-dimensional_2017} requiring feature sizes larger than the optical diffraction limit. Studying colloidal behavior in the limit of molecular-scale balances between attractive and repulsive interparticle interactions would require much smaller particle sizes and a different observation modality.

Although often intended as a means to passively image materials with high spatial resolution, electron irradiation often plays an active role in driving dynamic nanoscale or atomic processes. For instance, electron irradiation has been used to controllably etch nanoparticles,\cite{ye_single-particle_2016} induce crystallization in two-dimensional materials,\cite{bayer_atomic-scale_2018} drive particle assemblies at the nanoscale,\cite{kim_imaging_2017,luo_quantifying_2017} and manipulate the positions of individual atoms in a lattice \cite{su_engineering_2019}. Using electron irradiation for such processes provides high spatial and temporal resolution and precise control over electron-sample interactions by modulating the current and accelerating voltage of the electron beam. For recording or inducing nanoparticle dynamics, electron imaging allows for the precise mapping of particle trajectories leading to the formation of higher-order assemblies. Although in situ investigations of particle assemblies in liquid environments are becoming more prevalent,\cite{kim_visualizing_2016,kim_assessing_2019} the exact impact of electron irradiation on the dynamics of these assemblies is not well understood.

Merging high resolution and deliberate manipulation of colloidal systems with electron microscopy provides the opportunity both to explore interactions among smaller particles and to leverage and elucidate the influence of an electron beam on a colloidal system while employing it to drive different behaviors. Both the particle size and the electron beam enable interparticle interactions on different, shorter time scales that would otherwise be challenging to observe. While most colloidal studies are performed on monolayers in aqueous solutions, ionic liquids (IL)---replete with enigmatic properties in their own right---are interesting alternative media, especially for higher-resolution electron microscopy studies, owing to their low vapor pressures.\cite{torimoto_new_2010} Recent efforts have shown that colloidal nanoparticles can be incorporated into ILs\cite{zhang_stable_2017} or self-assembled at IL interfaces.\cite{frost_particle_2014} For example, Kim et al. recently showed that scanning electron microscopy (SEM) can image few-nanoparticle dynamics at the IL-vacuum interface and map interaction potentials between particles.\cite{kim_assessing_2019}

Here we present the dynamics of a 2D colloidal polycrystalline monolayer at an IL-vacuum interface driven by the irradiation of a focused, scanning electron beam. By introducing the combination of an IL interface and the charge-mediated actuation of the nanoparticles, we show that progressive charging of the 300-nm-diameter insulating silica nanoparticles by the electron beam causes a series of dynamic processes. First, particle charging, combined with compensation by IL screening, increases particle kinetic energy to mobilize the particles, resulting in annealing of the monolayer. Second, the particles detach from the IL-vacuum interface, resulting in the formation of point vacancies in the 2D monolayer, which leads to a melting phase transition. We first describe the dynamics that we induce and observe, then explore the macroscopic and microscopic dynamics of each stage, and finally describe a model to consistently explain the sum total of the observations. This newfound understanding of how the electron beam and IL cooperate to drive particle dynamics will enable a powerful new approach to manipulating colloidal systems without the need to vary their intrinsic properties.

\section{Results and Discussion}

\subsection{Driven colloidal monolayer dynamics}

\begin{figure*}[t]
 \centering
 \includegraphics[width=16.8cm]{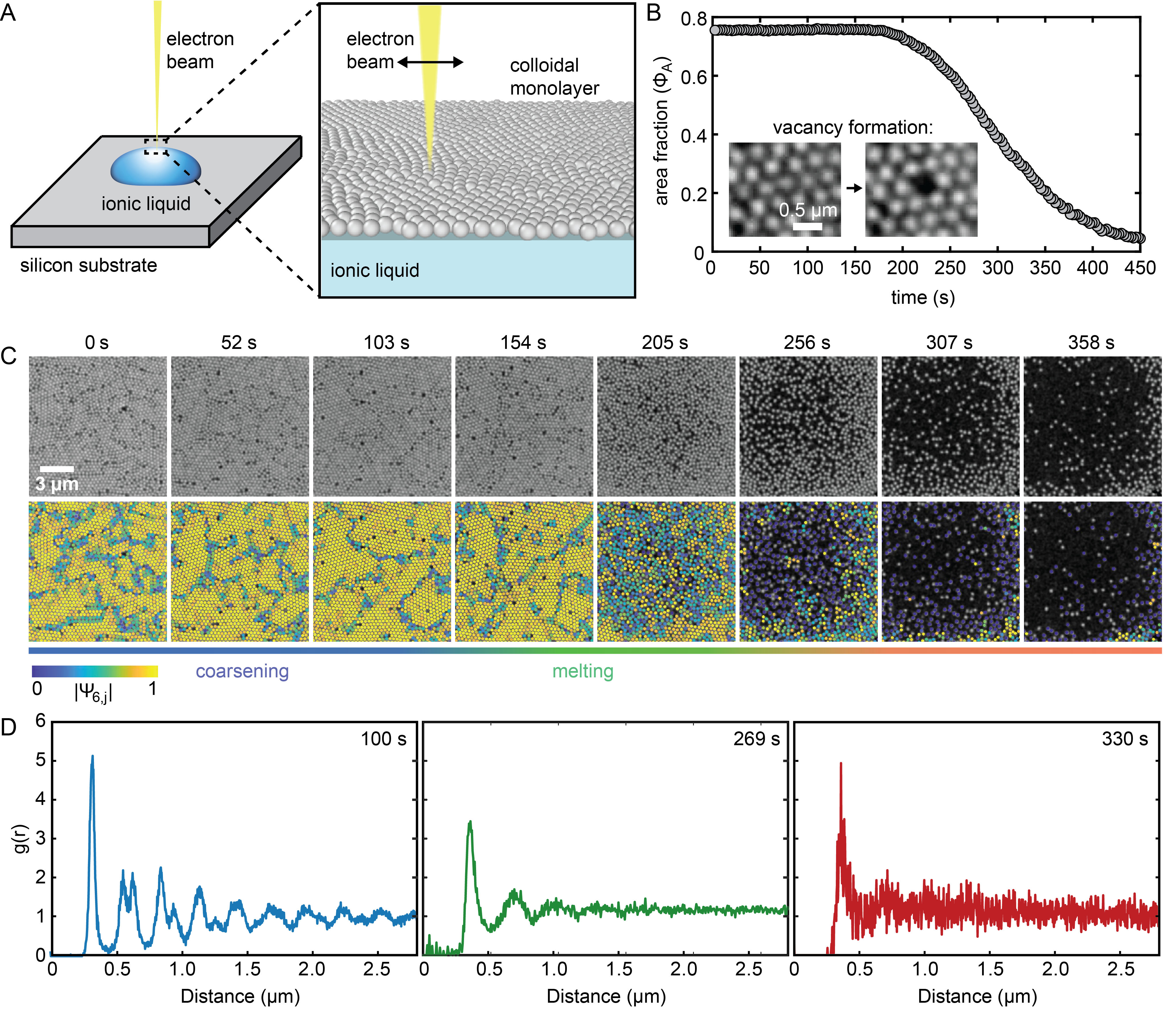}%.eps}
 \caption{Overview of coarsening, melting, and subsequent thinning upon electron beam irradiation. (A) Experimental setup showing the colloidal monolayer at the IL-vacuum interface and the electron beam scanning over the monolayer. (B) Area fraction of the monolayer as a function of time. The inset shows a single vacancy forming upon electron beam irradiation. (C) Time series of raw electron beam images and correlated image with false color indicating the magnitude of the single particle hexagonal bond order parameter ($\vert\Psi_6,_j\vert$) with increasing electron beam irradiation. It shows coarsening and then melting of the colloidal monolayer. (D) Radial distribution function $g(r)$ at 100 s, 222 s, and 330 s illustrates the progression from hexagonal solid to liquid to lower density liquid.}
\label{F1}
\end{figure*}

% Paragraph 1
Figure~\ref{F1}A shows the experimental setup, consisting of a polycrystalline colloidal monolayer that self-assembles at the interface between an IL droplet and vacuum. The self-assembled colloidal monolayer is formed by combining an aqueous solution of colloidal nanospheres (silica, 300 nm) with an IL (1-ethyl-3-methylimidazolium methyl sulfate). The mixture is pipetted onto a silicon substrate and placed in a vacuum chamber (40 mtorr). As the water evaporates, the beads assemble at the interface between the IL and vacuum environment. The polycrystalline monolayer is formed on a droplet of IL (1-3 mm diameter and $\sim$100-200 $\mu$m height) and placed inside a SEM. The monolayer, which covers a large region, primarily forms a hexagonal lattice although it contains many  point vacancies and disordered interfaces between crystalline grains (Figure S1). The formation of a polycrystalline monolayer suggests an attractive potential between the colloids. Upon repeated irradiation by a scanning, focused electron beam, the particles first begin to fluctuate, and then the monolayer begins to coarsen. Eventually, individual particles disappear from the interface, reducing the area fraction ($\phi_A$) until the monolayer melts (Figure~\ref{F1}B). The inset to Figure~\ref{F1}B shows a single particle disappearing from a crystalline region of the monolayer upon electron beam exposure. Most often, the particle completely disappears within a single frame, but in some cases, the particle disappears more gradually (Figure S2).

Figure~\ref{F1}C shows SEM images of the particle monolayer as a function of time, as well as corresponding images false colored by the magnitude of the single particle hexagonal bond order parameter ($\vert\Psi_6,_j\vert$). The images are acquired at an accelerating voltage of 5 kV, a beam current of 100 pA, and a scan rate of 2.56 s/frame. Movie S1 includes all recorded frames. Here, this bond order parameter is defined as:

\begin{equation}
\Psi_6,_j=\frac{1}{N_{nn}}\sum_{k=1}^{N_{nn}} e^{6i\theta_j,_k} \label{EQ1}
\end{equation}

\noindent where $N_{nn}$ is the number of neighboring particles within 0.4 $\mu$m, $\theta_j,_k$ is the angle between the bond vector connecting particle $j$ and neighboring particle $k$. $\Psi_6,_j$ quantifies the crystallinity of individual particles based on the spatial configuration of their nearest neighbors. The magnitude, ranging between 0 and 1, specifies the local order of the particles, while the phase provides the local orientation of a crystalline region that includes the particle, referenced to the horizontal direction in this case. We note that under this convention for $\Psi_6,_j$, a particle next to a vacancy that is otherwise configured with perfect hexagon symmetry with its five remaining neighbors retains a high magnitude of $\Psi_6,_j$. The false color aids in visualizing disordered regions (blue) and crystalline regions (yellow) of the film. Based on the time series of images, once the particles begin to fluctuate, we observe two primary regimes of dynamics. First, we observe that the number of blue disordered particles decreases with time, indicating coarsening of the monolayer. Second, after some time, individual particles begin to disappear, causing the area fraction to decrease and resulting in a higher fraction of particles with a low $\vert\Psi_6,_j\vert$ value (bluer), and ultimately to melting of the 2D solid, followed by conversion to a 2D vapor at low enough area fraction. These stages are all captured in $g(r)$, the radial distribution function of the monolayer as a function of time (Figure~\ref{F1}D), which describes the average spatial correlations between the colloidal particles and can be used to identify the structural phase \cite{chaikin_principles_1995}. At early times, $g(r)$ shows sharp peaks and long-range correlations characteristic of a solid hexagonal lattice (Figure~\ref{F1}D left). Over time, the peaks broaden and damp out to show a function representative of a liquid (Figure~\ref{F1}D middle). At the longest times, the disappearance of additional particles gives rise to a noisier liquid-like $g(r)$ profile  (Figure~\ref{F1}D right). Similar dynamics to those depicted in Figure~\ref{F1} have been measured reproducibly in separate trials, an example of which is shown in Figure S3. We next independently characterize the two primary regimes of particle dynamics, coarsening and melting, to better understand how the electron beam impacts particle motion in the self-assembled monolayer.

\begin{figure*}[t]
\includegraphics[width=17cm]{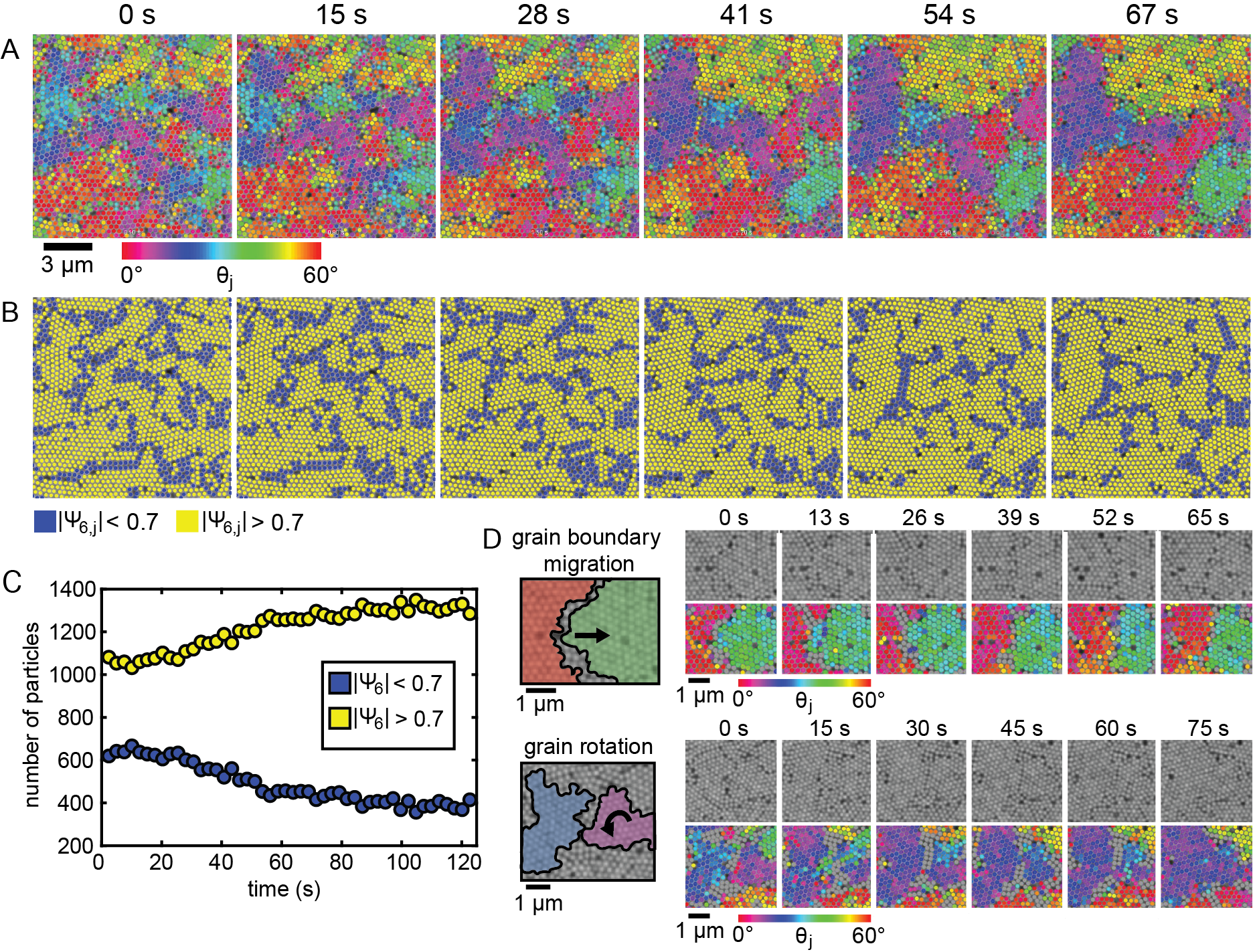}%.eps}
\centering
\caption{\label{fig2} Coarsening dynamics. (A) Time series of the colloidal monolayer false colored by the phase of the bond order parameter ($\theta_j$). (B) Number of disordered (blue, $\vert\Psi_6,_j\vert$ $<$ 0.7) and bulk crystalline (yellow, $\vert\Psi_6,_j\vert$ $>$ 0.7) particles as a function of time. (C) Average $\vert\Psi_6\vert$ as a function of time for crystalline (yellow, $\vert\Psi_6\vert$ $>$ 0.7) and disordered (blue, $\vert\Psi_6\vert$ $<$ 0.7) particles. (D) Two types of correlated motion of particles, grain interface migration and grain rotation, false colored by  $\theta_j$.}
\label{F2}
\end{figure*}

\subsubsection{Coarsening}
% Paragraph 2
Using single particle tracking, we obtain the dynamics of the coarsening at the level of the entire several-hundred nanoparticle field of view to elucidate how the degree of crystallinity increases over time, and we also examine smaller fields of view to follow individual events contributing to coarsening, such as grain merging and disordered interface area reduction. To visualize how the grain structure changes with electron beam irradiation, we calculate  $\theta_j$, the phase of $\Psi_6,_j$. The value of  $\theta_j$ describes the orientation of single particles relative to their neighbors and is a convenient way to visualize the evolution of grain structure. Figure~\ref{F2}A shows the evolution of the colloidal monolayer upon electron beam irradiation over the initial coarsening period. Initially at 0 s, the film is composed of many small crystalline grains, which merge into larger grains as the electron beam exposure time increases. This merging of grains occurs through a number of processes, including grain interface motion, grain rotations, and other more complex correlated particle motions. As the small grains merge, we observe a decrease in the number of disordered particles. To separately visualize disordered versus ordered particles, we false color particles with $\vert\Psi_6,_j\vert$ $<$ 0.7 in blue and particles with $\vert\Psi_6,_j\vert$ $>$ 0.7 in yellow (Figure~\ref{F2}B). We observe that the number of blue particles decreases substantially with time, indicating that disordered particles are being transferred to the crystalline phase. We also map rapid changes in particle crystallinity by plotting the change in $\vert\Psi_6,_j\vert$ from frame to frame ($\Delta\vert\Psi_6,_j\vert$), with $\Delta\vert\Psi_6,_j\vert$ $>$ 0.2 in purple and $\Delta\vert\Psi_6,_j\vert$ $<$ -0.2 in cyan overlayed on the SEM image (Figure S4). We find that the most substantial fluctuations occur at interfaces between grains. In Figure~\ref{F2}C we plot the number of crystalline ($\vert\Psi_6,_j\vert$ $>$ 0.7) and disordered ($\vert\Psi_6,_j\vert$ $<$ 0.7) particles. After about 20 s, the number of crystalline particles increases, while the number of disordered particles decreases. After approximately 80 s, both populations have nearly plateaued. Overall, we find that the electron beam anneals the particle monolayer over the first $\sim$100 s of imaging.

More microscopically, this coarsening occurs through both grain interface migration and larger-scale correlative motions, such as grain rotations. An example of each type of motion is included in Figure~\ref{F2}D. We show the SEM electron images and the corresponding color maps of  $\theta_j$. In the first case, the red grain takes over a larger portion of the field of view as the interface between it and the green domain migrates to the right. In the second case, the purple grain rotates until the two grains merge into a single grain (Figure S5A). At the single particle level, particles typically hop from one grain to the other to facilitate grain interface motion, yet in other cases, we find that sliding dislocations play a prevailing role. These sliding dislocations, shown schematically in Figure S5B, are observed often upon electron beam exposure. One function of these sliding dislocations is to shuttle point vacancies to disordered interfaces, as shown in Figure S5C. In this case, sliding dislocations serve as a mechanism to transport vacancies over relatively long distances from highly ordered regions to more disordered regions, such as grain interfaces. The shuttling of defects can increase the fluidity of the interfaces and contribute to their motions. Sliding dislocations also underlie grain rotations (Figure S5D). In Figure S5D, we observe a sliding dislocation forming in the middle of a grain. This sliding dislocation changes the overall grain orientation, which is also helped by relatively disordered regions surrounding the single crystal grain. The above examples emphasize the importance of correlated motions, such as sliding dislocations, of particles in the coarsening process. We next focus on the second regime of electron beam-induced particle dynamics, melting of the monolayer due to particle vacancy formation under electron beam irradiation.

\subsubsection{Melting}

\begin{figure*}[t]
\includegraphics[width=17cm]{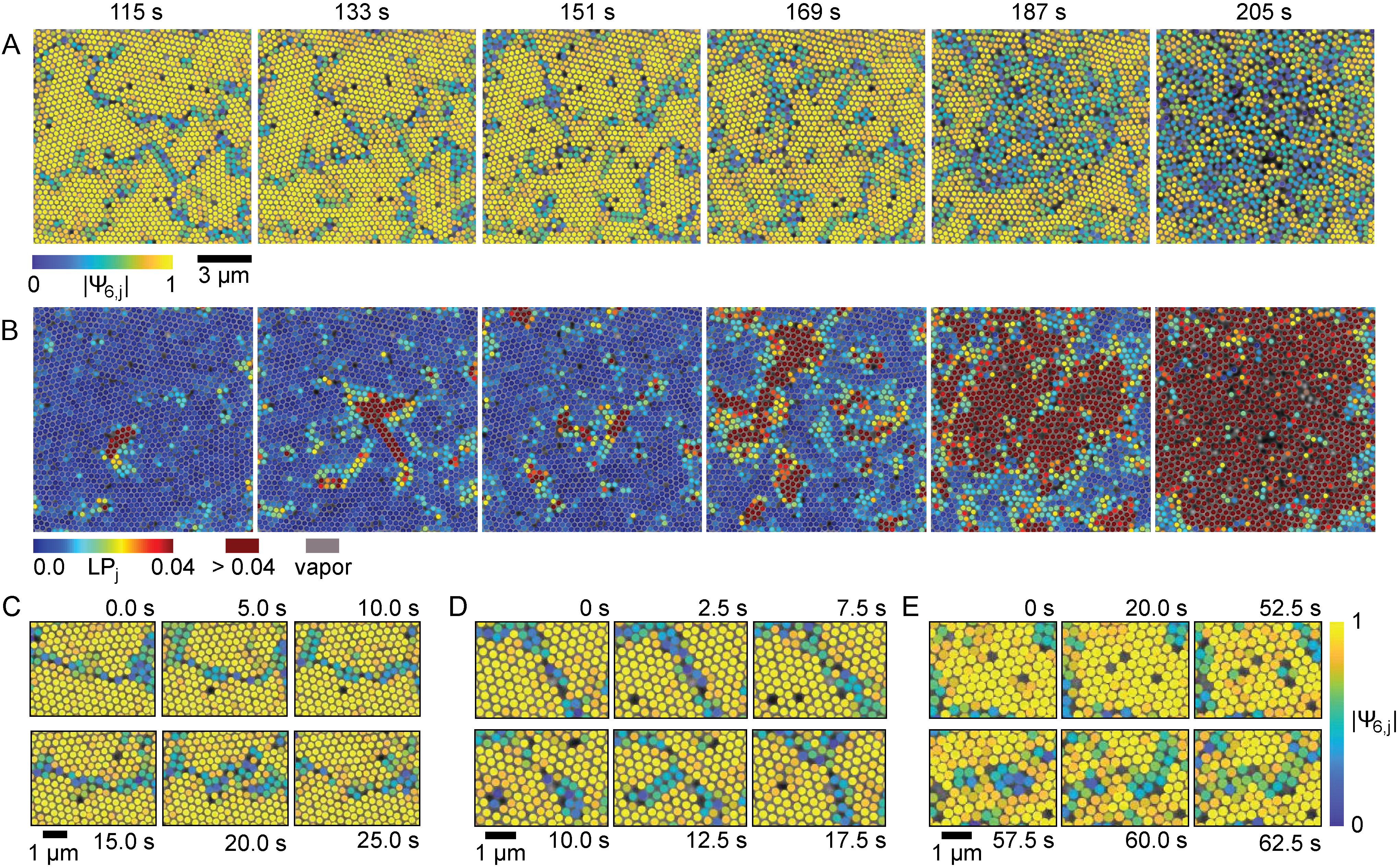}%.eps}
\centering
\caption{\label{fig3} Microscopic dynamics of melting and thinning. (A) Time series of particles false colored by the magnitude of the bond orientational order parameter ($\vert\Psi_6,_j\vert$). (B) Time series of particles false colored by the single particle instantaneous Lindemann parameter, $LP_j$, with vapor particles (no nearest neighbors) colored grey. (C) Example of a vacancy generated and then incorporated into a grain interface with false color showing $\vert\Psi_6,_j\vert$. (D) Example of two neighboring vacancies generated, merged, and then incorporated into a grain interface through a sliding dislocation with false color showing $\vert\Psi_6,_j\vert$. (E) Example of a series of generated vacancies forming a more disordered region with false color showing $\vert\Psi_6,_j\vert$.
}
\label{F3}
\end{figure*}

% Paragraph 4
We observe that a liquid phase emerges through the formation of vacancies in the crystal lattice. Figure~\ref{F3}A shows a time series of the particle monolayer after the initial coarsening occurs, between 115 and 205 s, false colored by $\vert\Psi_6,_j\vert$, as in Figure \ref{F1}C. As the electron beam scans, particles recede into the IL, effectively lowering the area fraction of the colloidal monolayer. Particles begin disappearing, and the area fraction decreases as shown in Figure~\ref{F1}B. We find that particles disappear both within crystalline and disordered regions. While particles are observed to disappear less frequently toward the edges of the field of view, this is only a boundary effect, with the irradiated particles being close to unirradiated ones beyond the field of view.

To gain further insights into the dynamics of the system upon particle deletion, we use single particle tracking and calculate an instantaneous 2D Lindemann parameter\cite{bedanov_modified_1985, zahn_two-stage_1999, kelleher_phase_2017, peng_two-step_2015, peng_diffusive_2017} for each particle $j$ as a function of time $t$ (Figure~\ref{F3}B), which we define as:

\begin{equation}
LP_j(\tau_0,t)= \frac{(\frac{1}{N_{nn}} \sum_{k}^{N_{nn}}(\Delta r_j(\tau_0,t)-\Delta r_k(\tau_0,t)))^2}{2a^2} \label{EQ2}
\end{equation}

\noindent where $\tau_0$ represents the time between successive frames, $N_{nn}$ is the number of nearest neighbors, $a$ corresponds to the average spacing between particles at time $t=0$, and $\Delta r_j(\tau_0,t)=r_j(\tau_0 + t)-r_j(t)$ indicates the displacement of particle  $j$ between successive frames. (See the Supplementary Information for details on how this expression is adapted from Ref.~\cite{kelleher_phase_2017}.) The instantaneous Lindemann parameter associated with a given particle $j$ captures the extent of correlated motions with its nearest neighbors, which therefore decreases upon melting.  We distinguish solid-like and liquid-like particles at any given time point by specifying a threshold value for the divergence of $LP_j$ of 0.04. Particles with $LP_j > 0.04$, colored in dark red in Figure~\ref{F3}B, are designated as liquid-like particles.\cite{peng_two-step_2015,peng_diffusive_2017} The liquid-like particles are initially, e.g. at 151 s, few in number. As more particles disappear, the 2D system enters a state of coexistence between a solid and liquid phase, with particles rapidly converting between solid and liquid states (e.g., 169 s). After enough particles disappear (e.g., 187 s), the liquid phase emerges.

Figure~\ref{F3}C-E show different microscopic mechanisms for the emergence of the liquid phase. In Figure~\ref{F3}C, a vacancy forms adjacent to a grain interface. As the interface moves, the vacancy is incorporated into it. Overall, the incorporation of vacancies leads to larger, more mobile disordered regions. In Figure~\ref{F3}D, two adjacent vacancies are formed. These vacancies combine and then are transported to more disordered regions through a sliding dislocation. These sliding dislocations, common motifs in the data, relay vacancies to disordered regions over a larger distance. Finally, Figure~\ref{F3}E shows how the formation of multiple vacancies can produce a new disordered region.

\begin{figure*}[t]
\includegraphics[width=17cm]{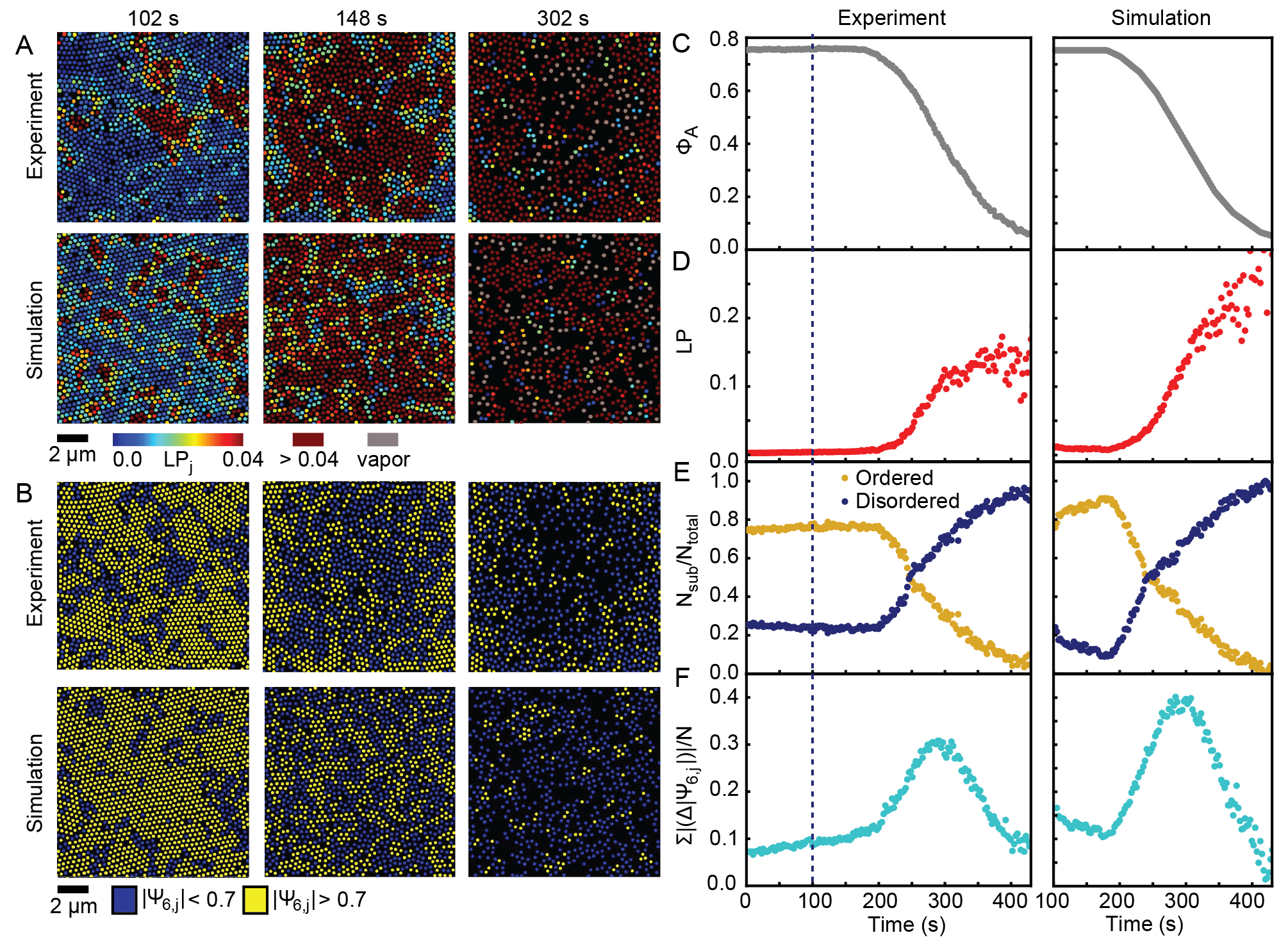}%.eps}
\centering
\caption{\label{fig4} Melting is consistent with simulations of monolayer dynamics. (A) Comparison of the single particle instantaneous Lindemann parameter $LP_j$ for the experiment and simulation as a function of time. (B) Comparison of $\vert\Psi_6,_j\vert$ as a function of time for experiment and simulation with $\vert\Psi_6,_j\vert$ $>$ 0.7 in yellow and $\vert\Psi_6,_j\vert$ $<$ 0.7 in blue. (C) Area fraction of particles as a function of time for experiment and simulation. Average instantaneous Lindemann parameter (LP) (D), fraction of particle with $\vert\Psi_6,_j\vert$ $>$ 0.7 in yellow and $\vert\Psi_6,_j\vert$ $<$ 0.7 in blue (E), and average  of frame-to-frame single particle bond order parameter fluctuations, $\sum \vert ( \Delta\vert\Psi_6,_j\vert ) \vert / N$, as a function of time (F) for both experiment and simulation. }
\label{F4}
\end{figure*}

%Paragraph 6:
Interestingly, the melting transition observed upon repeated electron beam irradiation can be recapitulated with simple Brownian dynamics simulations. Rather than explicitly including progressive particle charging, we need only use a Lennard-Jones interparticle potential with fixed parameters and a phenomenological, randomized particle deletion rate that follows the experimentally obtained area fraction versus time (Figure~\ref{F1}B). We carry out our simulations in reduced units and find that a temperature $T$ = 0.75 yields optimal qualitative agreement with the experiment, indicating that the system undergoes a solid to fluid melting transition upon progressive particle disappearance \cite{smit_vaporliquid_1998}. The simulation is initiated at a point corresponding to $\sim$100 s of evolution experimentally, indicated as a dashed vertical line in Figure~\ref{F4}C-F, by using the experimental configuration of particles at that point as the initial condition. Before this time point, the primary role of the electron beam is to increase particle fluctuations without decreasing the area fraction (see Movie S1).  Figure~\ref{F4}A compares 102, 148, and 302 s snapshots of experimental and simulation data false colored by the single particle instantaneous Lindemann parameter, $LP_j$. Figure~\ref{F4}B compares the same experimental and simulation frames, false colored instead by $\vert\Psi_6,_j\vert$, using the same false coloring scheme as in Figure~\ref{F2}B. These images show qualitative similarities between experiment and simulation.

To provide a more quantitative comparison, we evaluate a series of parameters while the area fraction of particles changes upon electron beam-induced particle disappearance for both the experiment and simulation (Figure~\ref{F4}C). These include the instantaneous Lindemann parameter of the entire system (Figure~\ref{F4}D), the number fraction of particles with $\vert\Psi_6,_j\vert$ both above and below a value of 0.7 (Figure~\ref{F4}E), and a representation of the average particle fluctuations (Figure~\ref{F4}F) via the average frame-to-frame change of the magnitude of the bond order parameter. A comparison of $g(r)$ at various times between simulation and experiment is shown in Figure S6. The good match in these parameters between the simulation and experiment suggests that the simulation captures the same dynamic processes as observed experimentally and that the experimental system is well described as a 2D Lennard-Jones system, which is known to undergo a density controlled melting phase transition \cite{smit_vaporliquid_1998}.

We first investigate the evolution of the system's instantaneous Lindemann parameter (Figure~\ref{F4}D). Upon a melting phase transition, the average value of the Lindemann parameter should diverge.\cite{bedanov_modified_1985} Through a similar trend in the experiment and simulation we find that the system's instantaneous Lindemann parameter begins to diverge shortly after the particles begin disappearing, at an area fraction still very close to its original value. This point in the evolution indicates the beginning of the melting transition and can be used as a reference point for other average properties. Next, we evaluate the number of non-crystalline (disordered) particles versus crystalline (ordered) particles as the system evolves using the same definition of crystalline versus noncrystalline particles as in Figure~\ref{F4}B ($\vert\Psi_6,_j\vert$ $<$ 0.7 are non-crystalline and  $\vert\Psi_6,_j\vert$ $>$ 0.7 are crystalline). Before the instantaneous Lindemann parameter begins to diverge, the number of crystalline and non-crystalline particles remains roughly constant (Figure~\ref{F4}E). After melting begins ($\sim$200 s), the number of crystalline particles sharply decreases, whereas the number of non-crystalline particles increases. This observation is consistent with an expansion of the disordered regions as particles are removed from the system. Last, we investigate the average frame-to-frame magnitude change in $\vert\Psi_6,_j\vert$, i.e., $\langle \vert ( \Delta\vert\Psi_6,_j\vert)\vert\rangle$, which shows how the magnitude of fluctuations between crystalline and non-crystalline states changes as particles disappear (Figure~\ref{F4}F). After melting begins, this ensemble averaged measure of fluctuations increases as more of the fluid phase emerges and then decreases as the fluid density continues to decrease. This steady increase occurs during coexistence between solid and fluid phases, in which particles are rapidly fluctuating between the two phases, and the rate is controlled by the rate at which the area fraction decreases. The subsequent decrease follows the further density reduction of the fluid phase. That the decrease begins before a plateau in $\langle\vert (\Delta\vert\Psi_6,_j\vert )\vert\rangle$ is obtained indicates that the area fraction is decreasing sufficiently quickly that the melting transition does not complete at a fixed fluid density because the decreasing area fraction protocol proceeds too quickly. The effects revealed by this ensemble averaged measure of fluctuations are also consistent between simulation and experiment. Together, these parameters demonstrate that the system undergoes a solid to fluid melting transition  upon progressive particle disappearance that involves the expansion of disordered regions and large fluctuations in crystallinity.

\subsection{Modeling observed behavior}
%Paragraph 7: describe the model that explains the observations
The experimental observations, taken together, raise the question of how the repeated scanning of the electron beam under unchanging conditions fosters a sequence of both initial mobilization and coarsening and also eventual melting and thinning. Furthermore, it is striking that a Lennard-Jones potential combined with the randomized particle deletion without explicit inclusion of beam interactions is able to recapitulate experimental dynamics, despite the intrinsic, perturbative role of the electron beam. We address these points with a self-consistent model. We first treat the progressively charged single particle interaction with the IL and vacuum and, second, address the use of the Lennard-Jones interparticle interaction.

\begin{figure}
\includegraphics[width=9cm]{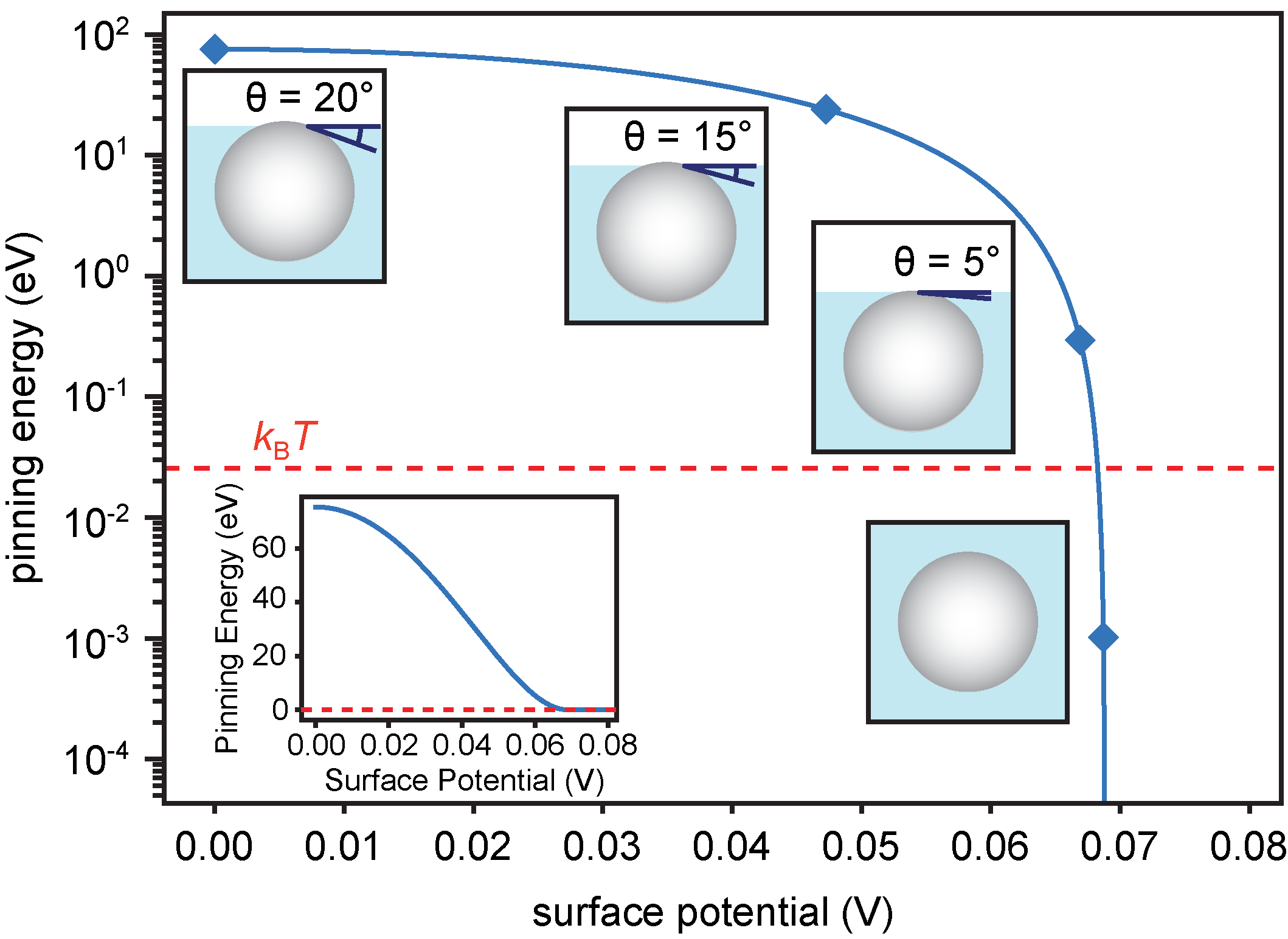}%.eps}
\centering
\caption{\label{fig5} Log plot of the interfacial pinning energy for a single particle as a function of surface potential. Schematics of the decreasing particle contact angle are shown at points along the curve. Particles thermally detach from the surface when the pinning energy is of order $k_{\rm B} T$ (red dashed line). Inset: Plot of pinning energy versus surface potential on linear axes.}
\label{F5}
\end{figure}

\subsubsection{Single particle interaction with the ionic liquid}
Particles initially decorate the IL-vacuum interface to minimize free energy.\cite{pieranski_two-dimensional_1980} At equilibrium, the energy balance for interfacial adsorption can be described in terms of the interfacial pinning energy, the work required to remove a single particle from the interface:\cite{williams_aggregation_1992,boker_self-assembly_2007}

\begin{equation}
\Delta E = \frac{\pi r^2}{\gamma_{L-V}}[\gamma_{L-S} - (\gamma_{V-S} - \gamma_{L-V})]^2 , \label{EQ3}
\end{equation}

\noindent where $r$ represents the radius of the particle cross-section at the interface, and the set of $\gamma_{I-J}$ refer to the surface tensions between the three different media---IL ($L$), vacuum ($V$), and solid particle ($S$) (Figure S7A). This pinning energy may be reexpressed in terms of the contact angle, $\theta$, using Young’s equation, which describes the contact angle strictly in terms of the three surface tensions, as:

\begin{equation}
\Delta E = \pi r^2\gamma_{L-V}(1-\cos\theta)^2 . \label{EQ4}
\end{equation}

\noindent The initial contact angle, calculated from known values for the media employed, and assuming zero charge on the silica particle, should be in the range of $\theta_{0}$ = 15-20$^{\circ}$.\cite{kim_assessing_2019} In this range, interfacial pinning energies are of order $10^3 \times$ the room temperature thermal energy ($k_{\rm B} T$); Only when the particles are nearly completely wetted with $\theta$ $<$ 2$^{\circ}$ does interfacial pinning become comparable to the thermal energy. We elaborate on this treatment in the Supplementary text and Figure S7A-B. SEM imaging causes each particle to charge over time, modifying the balance of interfacial interactions. In contrast with suspensions in polar solvents, where particle charging modulates electrostatic interparticle interactions,\cite{reincke_understanding_2006, luo_electrostatic_2012} in an IL particle charging primarily influences  screening by the solvent. More specifically, progressive particle charging stabilizes the particle-IL interaction, $\gamma_{L-S}$, and destabilizes the particle-vacuum interaction, $\gamma_{V-S}$, reducing the particle affinity for the interface and lowering the particle-IL contact angle. Equivalently viewed from the perspective of solvation, as a function of charging, more and more IL counterions must be recruited to the particle surface to screen a particle’s electric field, forming a capacitive double-layer.\cite{lynden-bell_electrode_2012} The counterions progressively cover the particle surface until none of the silica is exposed to vacuum, and the particle is no longer pinned to the interface. This electrowetting behavior as a function of the surface potential, $\psi$, is described by the Young-Lippmann equation:\cite{millefiorini_electrowetting_2006, horiuchi_calculation_2012}

\begin{equation}
\Delta\cos\theta=\frac{\epsilon_0\epsilon_r}{\lambda_D\gamma_{L-V}}\frac{(2k_{\rm B} T)^2}{e^2}(\cosh(\frac{e\psi}{k_{\rm B} T})-1) . \label{EQ5}
\end{equation}

\noindent Here, $\epsilon_0$ and $\epsilon_r$ are the vacuum permittivity constant and the IL dielectric constant, respectively, $\lambda_D$ is the IL Debye length, and $e$ is the electron charge. There is debate over the particular choice of Debye length, and we select a value of 3 $\rm \AA$ to reflect the capacitive screening length scale of the double-layer.\cite{weingartner_understanding_2008, perkin_self-assembly_2011, gebbie_long_2017} By parameterizing the contact angle in Equations \ref{EQ4} and \ref{EQ5}, we can directly assess the particle pinning energy as a function of surface potential as plotted in Figure~\ref{F5}. When the pinning energy is comparable to the thermal energy, the particle is able to desorb from the interface and be completely solvated by the IL. While the inset of Figure~\ref{F5} show the same model plotted on a linear scale, it is apparent on either plot that, at the point of unpinning, modest fluctuations of the surface pinning energy over a very narrow range in surface potential ($\sim$70 mV) are able to destabilize the pinning interaction. (See Table S1 for all model parameters used to obtain this estimate of the surface potential.) As the contact angle decreases, interfacial fluctuations enable more facile 2D translational motion and eventual unpinning.\cite{boniello_brownian_2015, hua_reversible_2016} Particle unpinning induces a solid-to-liquid phase transition in the 2D monolayer expressly by lowering the area fraction and corresponding 2D pressure. Given that the charging protocol for any given particle has a very low duty cycle while the electron beam scans the entire remainder of the field of view, we expect that the particles remain in equilibrium with their surrounding media though not necessarily with one another. In fact, the steeper changes in the simulation data seen between 100 and 200 s in the ordered and disordered particle fractions in Figure~\ref{F4}E likely arise because  the Lennard-Jones simulation is more readily able to equilibrate than the interparticle interactions allow for in the experiment.

%Paragraph 8: Describe suitability of interparticle model
\subsubsection{Interparticle interactions}
Having provided an explanation for particle unpinning, we turn to describing the effect of progressive particle charging on interparticle interactions within the monolayer at the interface and further assess the validity and limitations of employing the Lennard Jones model. We suspect that Coulomb interactions are well screened by the IL and that they do not directly influence interparticle interactions.\cite{ueno_colloidal_2008,he_nanoparticles_2015} This assumption is supported by our determination that a  Lennard-Jones potential with fixed parameters is able to recapitulate the interparticle dynamics in the monolayer once they are set in motion and while they continue to be further charged by the electron beam. This recapitulation is illustrated by the good agreement in Figure~\ref{F4}  between experimental and simulation data. The exception noted above--that before any substantial decrease in area fraction the experimental system is not at equilibrium--indicates that additional interparticle interactions or particle-interface interactions likely do exist./cite{structured-colloids} The recovery of experiment-model agreement once the area fraction more precipitously decreases suggests that the area fraction changes dominate the simulation dynamics during this stage. To gain insight into the physical origin of the interaction between particles within the monolayer, we varied the interaction potential. Using a Weeks-Chandler-Anderson (WCA) potential \cite{weeks_role_1971} in order to eliminate attractive interparticle interactions and approximate a hard spheres model yielded generally faster  dynamics and qualitative disagreement with experimental data (Figure S8). Since our model for particle unpinning suggests that the IL effectively screens the progressively increasing charge on the particles, the interparticle potential appears to be rather invariant to charge build-up from repeated irradiation after the initial stage in which the particles are made to fluctuate. The exact nature of the interparticle forces presents an exciting avenue of future research. The strong qualitative agreement of the Lennard-Jones simulation with the experimental data suggests that van der Waals forces may be at play. Prior work, however, suggests that lateral capillary forces may also be present in our system.\cite{kralchevsky_capillary_2000, nikolaides_electric-field-induced_2002,oettel_effective_2005} This effective attractive interaction would be mediated by a combination of charging and the dielectric mismatch across the IL-vacuum interface.\cite{kaz_physical_2012}

\subsection{Control over the particle charging rate}

\begin{figure*}[t]
\includegraphics[width=17cm]{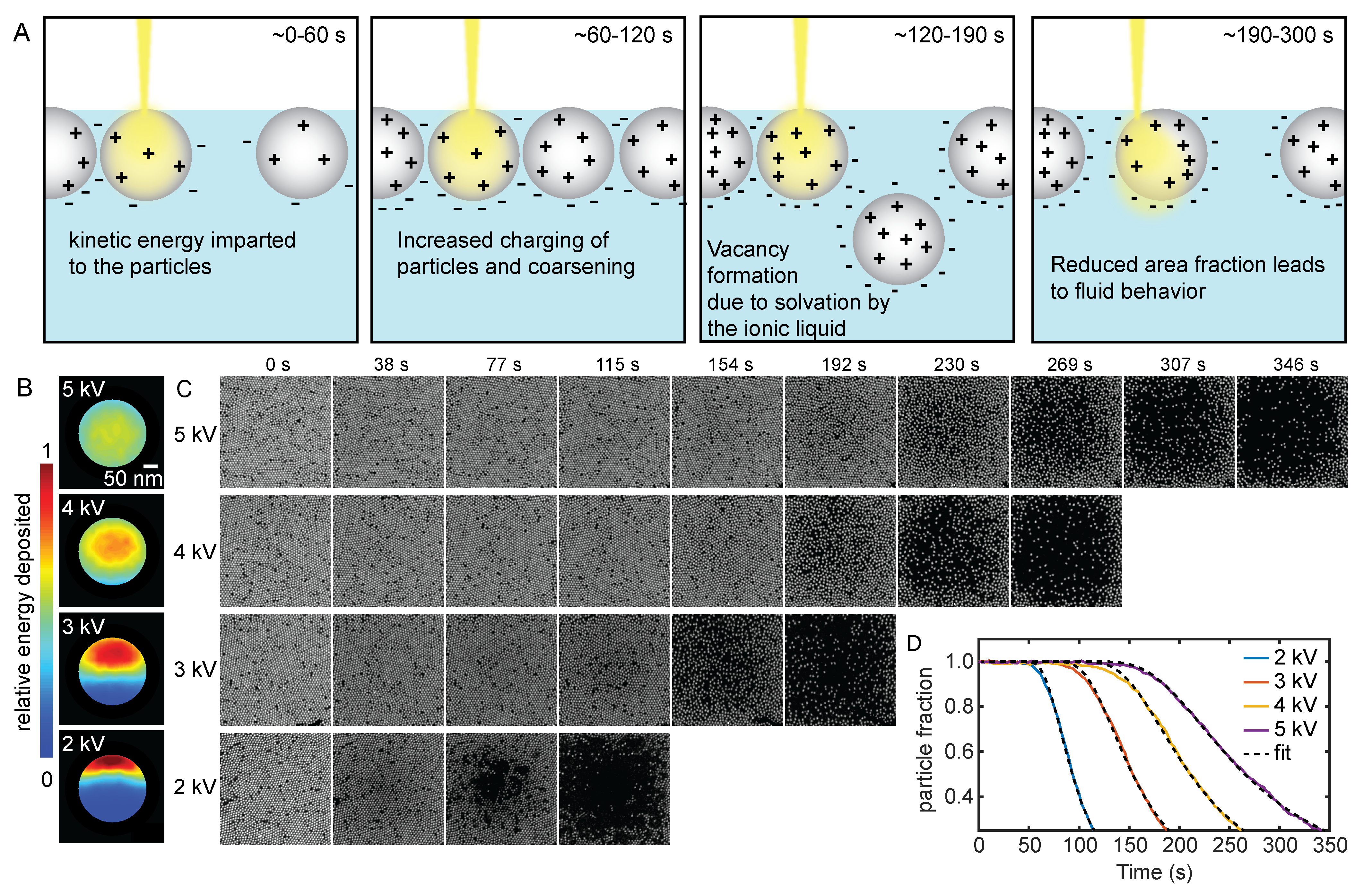}%.eps}
\centering
\caption{\label{fig6} Mechanism of particle coarsening and sinking and accelerating voltage dependence. (A) Schematic showing how the electron beam charges the silica particles to induce coarsening and then sinking into the IL layer. While the contact angle decreases with charging, it is not exaggerated, in order to most accurately depict the true scenario. (B) Simulated energy deposition of primarily electrons in a 300 nm silica sphere by electron beam as a function of accelerating voltage. The same energy scale is employed for all four cases to facilitate direct comparison. (C) Time series of SEM images showing the accelerating voltage dependence of coarsening and sinking. (D) Particle fraction as a function of time for 2-5 kV.}
\label{F6}
\end{figure*}

%Paragraph 9: casino simulations of kV-dependence
The model that we have established to explain the observed behavior of the electron beam-induced monolayer evolution, diagrammed in Figure~\ref{F6}A, is further supported by additional studies of the same monolayer system performed at progressively lower accelerating voltages. As a prelude to these experiments we describe the anticipated effect of lowering the accelerating voltage from the 5 kV that is used in all previously described experiments. Standard Monte Carlo simulations\cite{demers_three-dimensional_2011} of our electron beam irradiation of a silica particle at 2, 3, 4, and 5 kV yield the distributions of energy deposited in the particle shown in Figure~\ref{F6}B. The same scale is used in all four simulation results. We take the energy deposited to correlate to the inelastic scattering events of the beam’s electrons in the particle, which should also be a good proxy for the extent of charging. Altogether, particle charging reflects the balance of trapped primary electrons and of holes generated concomitantly with secondary electrons  (Figure S9). Because they are insulating we expect the charge distribution created in the particles by the electron beam to persist because there is neither a mechanism to transport charge within the particle nor to carry the charge out of the particle. The simulation results show that at 5 kV, the highest accelerating voltage explored, energy is deposited relatively uniformly over an entire particle. Decreasing the beam voltage decreases the size of the scattering volume to the point where at 2 kV it is concentrated near the surface in the top third of the particle. Additionally, because the scattering volume is largely  confined to the particle, relatively few beam electrons are transmitted through the particle, leading to higher rates of energy deposition and charging. Although the panels in Figure \ref{F6}B show asymmetric energy distributions, we suspect that repeated irradiation might distribute charge in the particle with spherical symmetry if the particle were able to fluctuate and hence rotate or revolve in place. This action, which could occur at other accelerating voltages as well, would serve to minimize the free energy of the particle by allowing the charge initially deposited at the vacuum interface to be best solvated by the IL.

Returning to the experimental data, Figure~\ref{F6}C shows the corresponding time series of SEM images at 2 kV, 3 kV, 4 kV, and 5 kV, keeping the current constant at $\sim$100 pA and the scanning time per frame constant at 2.6 s. In all cases, the particles ultimately disappear into the IL. The particles disappear at an earlier time at lower accelerating voltages than at higher accelerating voltages. The solid curves in Figure~\ref{F6}D shows the fraction of particles remaining at the interface as a function of time for 2-5 kV. Not only do the particles disappear from the monolayer sooner at lower accelerating voltages, but the distribution of disappearance times is also narrower (Figure S10). The median sinking time ($t_s$ in Table S2) in Figure \ref{F6}D scales roughly linearly with accelerating voltage, starting from $\sim$93 s at 2 kV and going up to $\sim$265 s at 5 kV. The unpinning model proposed above is consistent with this trend in accelerating voltage, presumably due to a fairly linear relationship between the surface potential that results from a particular accelerating voltage and the time taken for each particle to achieve that potential. We expect that the lower electron transmission, and consequently higher charging rate, at lower accelerating voltage increases the surface potential to recruit more countercharge from the IL and to lower the contact angle more quickly, promoting unpinning of the particles from the interface. Interestingly, the number of particles remaining at the interface as a function of time in addition to its accelerating voltage dependence can be captured with a phenomenological model in which particle charge is normally distributed with a standard deviation that linearly increases as a function of time (dotted curves in Figure~\ref{F6}D). The trajectories measured at the higher accelerating voltages that have later average disappearance times feature intrinsically broader distributions, as variation in the charge increases over time. Discussion of the fit is elaborated in the Supplementary Information (see fit parameters in Table S2). The finite slope of the curves indicates a broad distribution of unpinning times, which we attribute to variations in charging rate associated with local particle environment, the effective dose across the imaging field of view, and variations in electron beam exposure as particles fluctuate at the liquid-vacuum interface. Beyond these hypotheses to explain the particle disappearance profiles, the accelerating voltage dependence shows that changes to the properties of the electron beam can have a dramatic change on the behavior of the self-assembled particles. The longer wait time before particles begin to disappear at higher accelerating voltages allows more time to first reorganize, such that ordering is more substantial than at lower accelerating voltages.

\section{Conclusion}

We have examined and elucidated the mechanisms for charging-driven coarsening and melting of an initially static nanoparticle colloidal monolayer at an IL-vacuum interface by prompting and probing with a scanning electron beam. The coarsening of the initially defected 2D hexagonal lattice occurs through grain interface migration and larger-scale motions such as grain rotations, which are both often underlied more microscopically by sliding dislocations. The melting transition appears to be driven by a progressive decrease in the area fraction of the monolayer and is well-described by  Lennard-Jones interparticle interactions, combined with a phenomenological protocol for particle deletion. We have found that particle disappearance from the monolayer, which is at the heart of the decreasing area fraction that drives the monolayer phase transition, can be explained by considering the contact angle and associated pinning energy of a particle at the interface as a function of a surface potential that increases with increased charging by the electron beam. Although the particles are initially strongly pinned at the interface, charging recruits progressively more counterions from the IL, decreasing the pinning energy. Full wetting occurs when the particle surface potential is substantial enough to weaken the pinning interaction to the point where it is comparable to thermal fluctuations. Exploring the dynamics as a function of accelerating voltage and concomitant particle charge density profiles further supports our model of the phenomenon by showing how the characteristic time scales in the dynamics can be varied as a function of charging rate. We have thus elucidated a new mechanism by which to externally drive a colloidal system, with the exciting possibility to control behaviors without having to change the intrinsic properties of the particles being studied.

Looking forward, combining the independently rich parameter spaces of colloidal particles, ILs, and electron beam attributes stands to generate a wide range of new driven, emergent phenomena. First, in addition to exploring particle shapes, symmetries, and sizes, different electronic properties should strongly impact their interaction with the IL, with the electron beam, and therefore also with one another. For example, Janus-like insulating/conducting particles and other anisotropic forms should yield orientation-dependent, bond-like interactions, more closely approximating molecular systems, and should access new interaction regimes, some of which will be strongly dominated by electrostatics.\cite{zhang_janus_2017}
Second, since IL properties strongly affect the tendency toward monolayer assembly, exploring their composition could present an interesting platform to probe the frontiers of lateral capillary forces, as a high monolayer density could lead to many body effects theorized to strongly enhance capillary forces.\cite{pergamenshchik_strong_2009} Third, manipulating electron beam current, accelerating voltage, and X-Y scanning trajectories could readily be used to controllably perturb colloidal systems in more sophisticated ways as well, including tuning the accelerating voltage to preserve particle charge neutrality. This control could prepare initial or dynamically modulated conditions with specific surface pressures by corralling particles, could be used to selectively remove particles, by analogy with optical tweezers, and could generate more sophisticated time-dependent interparticle interactions. Finally, further decreasing particle sizes beneath what are typically employed in colloidal experiments while interrogating them non-perturbatively is still important to pursue in order to more closely approximate the interactions of molecular or atomic systems and to access shorter dynamical time scales. Developing other schemes that leverage electron microscopy resolution but that can non-invasively study more nanoscopic systems will be a valuable complement to electron beam-induced characterizations.\cite{bischak_cathodoluminescence-activated_2015,bischak_noninvasive_2017} Non-invasive imaging approaches offer the chance to compare the dynamics of a given nanoparticle system in the absence of charging to the same system upon direct electron beam irradiation.

\section*{Experimental}

\subsection*{Colloid Monolayer Preparation}
The colloid monolayer was prepared by mixing 20 $\mu$L stock colloid solution (300 nm diameter 10 mg/mL in water, nanocomposix) with 20 $\mu$L 1-ethyl-3-methylimidazolium ethyl sulfate (EMIM CH$_3$CH$_2$SO$_4$, Aldrich). 10 $\mu$L of the solution was then deposited on a $\sim$$1 \times 1$ cm Si substrate (Virginia Semiconductor), which was first cleaned by sonication in isopropyl alcohol (Aldrich) and acetone (Aldrich) for 5 min each. The sample was then placed in a vacuum chamber at 40 mtorr for 20 min to evaporate the water.

\subsection*{Scanning Electron Microscopy}
Colloidal monolayers on Si were grounded with a copper clip and loaded into a Zeiss Gemini SUPRA 55 S2 scanning electron microscope (SEM). Movies of electron beam-induced colloid motion were acquired using the in-lens detector with a 5 kV accelerating voltage (unless otherwise specified), a beam current of $\sim$100 pA, a scan rate of 0.01 s/line, and a field-of-view of approximately $12 \times 12$ $\mu$m (8000$\times$ magnification, $256 \times 256$ pixels). Movies of colloid dynamics were acquired using custom python software\cite{bischak_heterogeneous_2015} and saved as HDF5 files.

\subsection*{Single Particle Tracking}
Single particle tracking of the electron beam-induced colloidal monolayer dynamics was performed using the TrackMate plugin\cite{tinevez_trackmate_2017} of ImageJ.\cite{schindelin_fiji_2012} HDF5 files were imported using the HDF5 Plugin for ImageJ and Fiji. The movies were binarized using the Otsu method and then loaded into TrackMate. The parameters for the single particle tracking using TrackMate were a blob diameter of 0.3 $\mu$m, a threshold of 0.1, a maximum linking distance of 0.250 $\mu$m, a gap-closing maximum distance of 0.1 $\mu$m, and a gap-closing frame gap of 1.

\subsection*{Analysis of Single Particle Tracking Outputs}
Output files from TrackMate were loaded into a custom Matlab code. The custom Matlab code was used to compute the bond order parameter magnitude ($\vert\Psi_6,_j\vert$), bond order parameter phase ($\theta_j$), and instantaneous 2D Lindemann parameter ($LP_j$) for each particle $j$. Movies and times series of frames were generated with false colors that indicate $\vert\Psi_6,_j\vert$, $\theta_j$, $LP_j$, and $\Delta\vert\Psi_6,_j\vert$. Disordered and crystalline particles were distinguished by $\vert\Psi_6,_j\vert$ $<$ 0.7 and $\vert\Psi_6,_j\vert$ $>$ 0.7, respectively, and plotted as blue and yellow false-colored particles, respectively. A threshold value of 0.7 was chosen to distinguish crystalline and disordered particles by plotting the distribution of $\vert\Psi_6,_j\vert$ for the initial frames and finding the midpoint between the two subpopulations.

\subsection*{Monte Carlo Simulations of Electron Trajectories}
Monte Carlo simulations of electron energy deposition were performed using the CASINO v3.2 program.\cite{demers_three-dimensional_2011} For each simulation, the 3D trajectories of 10,000 electrons were simulated. Simulations were carried out at accelerating voltages ranging from 2.0 kV to 5.0 kV. For each accelerating voltage, energy deposited in a 300 nm-diameter sphere with a density of $\sim$2.5 g/cm$^3$ was calculated and integrated along one axis in order to represent the results in Figures \ref{F6}B and S9.

\subsection*{Molecular Dynamics Simulations}

The simulations presented in the main text were performed using LAMMPS software.\cite{plimpton_fast_1995} The initial conditions were extracted from the experimental data. Each simulation was carried out in two dimensions, and periodic boundary conditions were used. All simulations were carried out in reduced Lennard-Jones units with a fundamental unit of time given by $ \tau_{\rm LJ} = \frac{\sigma^2}{D} $, where $ \sigma $ and $ D $ are given in non-reduced units, and are the diameter and diffusion coefficient for an isolated colloidal particle, respectively. The diffusion coefficient was calculated using the Stokes-Einstein relation for a particle in an ionic-liquid (IL) solution \cite{requejo_effect_2014}. Each simulation is carried out with a time step of $ \delta t = 0.001 $, and we use a Langevin thermostat with a relaxation time of $ 200 \delta t $. Each simulation was initialized by using the experimental data frame at 100 seconds of experimental time, as this is the time required for the electron beam to drive the colloidal particles into a well defined state of motion.

\section*{Acknowledgments}

We thank D. T. Limmer and K. K. Mandadapu for insightful discussions and input. We thank E. Wong, E. S. Barnard, D. F. Ogletree, and S. Aloni at the Molecular Foundry for assistance with SEM. This work has been supported by STROBE, A National Science Foundation Science \& Technology Center under Grant No. DMR 1548924. The SEM imaging at the Lawrence Berkeley Lab Molecular Foundry was performed as part of the Molecular Foundry user program, supported by the Office of Science, Office of Basic Energy Sciences, of the U.S. Department of Energy under Contract No. DE-AC02-05CH11231. C.G.B. acknowledges an NSF Graduate Research Fellowship (No. DGE1106400), and N.S.G. acknowledges an Alfred P. Sloan Research Fellowship, a David and Lucile Packard Foundation Fellowship for Science and Engineering, and a Camille and Henry Dreyfus Teacher-Scholar Award.

\section*{Author Contributions}

C.G.B, J.G.R., J.W.K. and N.S.G. wrote the manuscript. C.G.B. and N.S.G. conceptualized the experiment. C.G.B. performed the electron microscopy experiments and single particle tracking analysis. C.G.B. and J.W.K. performed the molecular dynamics simulations. J.G.R. developed the unpinning model of the electron beam-irradiated colloids.\\

\section*{References}

%\bibliography{maintext_bib.bib}
\bibliography{maintext_bib}

\end{document}